\documentclass[aps,pra,twocolumn,showpacs,superscriptaddress,groupedaddress]{revtex4}
\usepackage{graphicx}
\usepackage{dcolumn}
\usepackage{bm}
\usepackage{color}
\usepackage{amsthm}
\usepackage{amsmath,amssymb}
\usepackage{amsfonts}

\usepackage[latin9]{inputenc}
\usepackage{esint}
\usepackage{bbm}
\usepackage{amsfonts}
\usepackage{mathrsfs}

\begin{document}
\title{Quantum parameter estimation via dispersive measurement in circuit QED}
\author{Beili Gong}
\author{Yang Yang}
\author{Wei Cui}\email{aucuiwei@scut.edu.cn}

\address{School of Automation Science and Engineering, South China University of Technology, Guangzhou 510641, China}
\date{\today}

\begin{abstract}
We investigate the quantum parameter estimation in circuit quantum electrodynamics via dispersive measurement.
Based on the Metropolis Hastings (MH) algorithm and the Markov chain Monte Carlo (MCMC) integration, a new algorithm is proposed to calculate the Fisher information by the stochastic master equation for unknown parameter estimation.
Here, the Fisher information is expressed in the form of log-likehood functions and further approximated by the MCMC integration. Numerical results demonstrate that the single evolution of the Fisher information can probably approach the quantum Fisher information. The same phenomenon is observed in the ensemble evolution in the short time interval. 
These results demonstrate the effectiveness of the proposed algorithm.
\end{abstract}
\maketitle

\section{Introduction}
The problem of accurately estimating unknown parameters in quantum system is of both fundamental and practical importance.
According to the parameter estimation theory \cite{Helstrom1976,Holevo2011,Wiseman2010,Escher2011},
in classical system the estimation precision is limited by the standard quantum limit (SQL) \cite{giovannetti:2004}, $1/\sqrt N$, where $N$ refers to the number of experiments.
In quantum system, Refs.~\cite{caves:1980,caves:1981} showed that with the help of squeezed state technique the parameter estimation accuracy can exceed the SQL, and even approach the Heisenberg limit (HL)\cite{zwierz:2010}, $1/ N$.
The classical Fisher information (FI) is a tool widely used to calculate the parameter estimation accuracy, and the Cram\'{e}r-Rao bound states that the inverse of the Fisher information is a tight lower bound on the variance of any unbiased estimation parameter \cite{hodges2012cramer,Zhong:2013}.
By explicitly maximizing the Fisher information over all possible measurement strategies, one can obtain the quantum Fisher information (QFI) \cite{Wang:2016,Li:2013,Zhang:2013,Jacobs2014Quantum}.

Over the past decades, parameter estimation via continuous weak measurement in quantum system caused a wide range of interests \cite{aharonov1988result,Smith:2006,Jacobs2014Quantum,zhang2015precision,Xu:2013,Ralph2011}.
Ref.~\cite{zhang2015precision} showed that weak measurements have a rich structure, based on which more novel strategies for quantum-enhanced parameter estimation can be constructed.
Ref.~\cite{Xu:2013} experimentally demonstrated a new robust method for precision phase estimation based on quantum weak measurement. \cite{Breuer:2002}
The stochastic master equation with quantum weak measurement was also derived for quantum parameter estimation \cite{Ralph2011}.
Moreover, the likelihood function and the statistical properties of the measurement output were demonstrated to be effective resources for quantum parameter estimation \cite{Gammelmark2013,Gammelmark2014}.
Although much progress has been made in quantum parameter estimation based on continuous weak measurement, 
 how to effectively calculate the Fisher information (or the estimation precision)  based on these resources is still 
 with remarkable difficulty. To figure out this problem, one needs to represent the Fisher information  in computable forms and take effective measures to prior-estimate the parameter of interest. 
A preliminary work \cite{Genoni:2017} to calculate the Fisher information based on various weak measurements for linear Gaussian quantum system was reported recently.
In this paper, we propose an efficient algorithm to calculate the Fisher information based on the quantum stochastic master equation in circuit quantum electrodynamics (circuit QED) \cite{You:2005,You:2011}.

Circuit QED is widely regarded as an excellent platform for quantum estimation and quantum control \cite{Wiseman2010,Blais:2004,Wallraff:2004,Chiorescu:2004,Vijay:2012,Slichte:2012,Xiang:2013,Cui:2013}.
Dispersive measurement in circuit-QED leads to a diffusion like evolution for the system and the measurement record, including the homodyne gain and the innovation. Due to the randomness of the measurement record, the numerical differentiation approach is used to calculate the derivative of the log-likelihood function, and a series of parameters of interest is randomly generated  by the Metropolis Hastings (MH) algorithm \cite{Gilks1996Introducing}.  Finally, the calculable Fisher information is approximated by the  Markov chain Monte Carlo (MCMC) integration \cite{Johnston2014Efficient,Efendiev:2008}.

This paper is organized as follows. In Sec.~\ref{Quantum_Parameter_Estimation}, a brief introduction of quantum parameter estimation is presented. In Sec.~\ref{Quantum_Weak_Measurement}, we discuss the dispersive measurement in circuit QED. The reduced stochastic master equation and the measurement record are exhibited in this section. An efficient algorithm to calculate the Fisher information is introduced in Sec.~\ref{Fisher_Information}.
Numerical experiment in circuit-QED demonstrates the feasibility and effectiveness of the proposed algorithm.
We summarize our conclusion in Sec.~\ref{Conclusion}.

\section{Quantum Parameter Estimation}\label{Quantum_Parameter_Estimation}
Suppose $\theta$ is an unknown parameter that needs to be estimated in a quantum system. As we mentioned above, the precision of the unbiased parameter estimation is always indicated by the quantum Cram\'{e}r-Rao inequality \cite{hodges2012cramer,Braunstein1994Statistical,Ciampini:2016}, i.e.,
\begin{equation}\label{CR_bound}
 \left\langle {{{\left( {\delta \theta } \right)}^{\rm{2}}}} \right\rangle  \ge \frac{1}{{NI\left( \theta  \right)}},
\end{equation}
where $I\left( \theta  \right)$ is the Fisher information of $\theta$, ${\delta {\theta}}$ is  the estimation error,   and $N$ is the number of measurements.

Let $D$ be the measurement output, which is conditioned on the value of the unknown parameter $\theta$.
The ability to estimate the unknown parameters depends on the probability of observing the output given the parameters $P(D|\theta)$, which can be characterized by the Fisher information, i.e.,
\begin{equation}\label{FI_0}
  I\left( \theta  \right) = E\left[ {{{\left( {\frac{{\partial \ln P\left( {D\left| \theta  \right.} \right)}}{{\partial \theta }}} \right)}^2}} \right],
\end{equation}
where $E\left[  \cdot  \right]$ refers to the expectation value with respect to independent realizations of the measurement results $D$.
Sometimes the probability density $P (D|\theta)$ is also defined as a likelihood function.
In addition, the theory that tackles the probability distribution of the measurement resource is the same as for the classical problems with stochastic measurement outcomes while the underlying dynamics of the system and $P(D|\theta)$ may be dominated by the laws of quantum physics \cite{Kiilerich2016}.

By maximizing $I\left( \theta  \right)$ over all possible quantum measurements on the system, one can obtain the quantum Fisher information (QFI) \cite{rossi2016enhanced}. Simply, if a quantum pure state ${\rho _{\theta}} = \left| {{\psi _{\theta}}} \right\rangle \left\langle {{\psi _{\theta}}} \right|$  evolves in a closed quantum system, the quantum Fisher information of the parameter is given by
 \begin{equation}\label{FI_1}
 I = 4\left[ {\left\langle {{\psi_{\theta} '}}
 \mathrel{\left | {\vphantom {{\psi_{\theta} '} {\psi_{\theta} '}}}
 \right. \kern-\nulldelimiterspace}
 {{\psi_{\theta} '}} \right\rangle  - {{\left| {\left\langle {\psi_{\theta} }
 \mathrel{\left | {\vphantom {\psi_{\theta}  {\psi_{\theta} '}}}
 \right. \kern-\nulldelimiterspace}
 {{\psi_{\theta} '}} \right\rangle } \right|}^2}} \right],
\end{equation}
where $\left| {\psi_{\theta} '} \right\rangle $ stands the derivative of $\left| {\psi_{\theta}} \right\rangle$ with respect to the parameter $\theta$.

\section{Dispersive Measurement in Circuit QED}\label{Quantum_Weak_Measurement}
Circuit QED consists of a superconducting qubit and a microwave resonator cavity.
The superconducting system can be described by a two-level quantum system with the Hamiltonian
\begin{equation}\label{H}
H =  \frac{\Delta }{2}{\sigma_x}+\frac{\Omega}{2} {\sigma _z},
\end{equation}
where $\Delta$ is the electrostatic energy and $\Omega$ is the Josephson energy \cite{You:2005,Wallraff:2004,Devoret:2013}.

By applying a displacement transformation and tracing over the resonator state, we can eliminate the cavity degrees of freedom and get a reduced stochastic master equation \cite{Gambetta:2008,Qi:2010,Feng:2016} with dispersive measurement $(\hbar=1)$
\begin{equation}\label{Eqe:stochastic_master_equation_1}
\begin{aligned}
d{\tilde \rho _t} =  - i\left[ {H,{{\tilde \rho }_t}} \right]dt + \eta {\cal D}\left[ F \right]{\tilde \rho _t}dt + \sqrt \eta  \mathcal{M}\left( {{{\tilde \rho }_t}} \right)d{Y_t},
\end{aligned}
\end{equation}
with
\begin{equation}\nonumber
\begin{aligned}
\mathcal{D}\left[ A \right]\rho  &= A\rho {A^\dag } - \frac{1}{2}\left( {{A^\dag }A\rho  + \rho {A^\dag }A} \right);\\
\mathcal{M}\left( \rho  \right) &= A\rho  + \rho {A^\dag }.
\end{aligned}
\end{equation}
Here $\tilde{\rho_t}$ is the un-normalised state, $F$ is the measurement operator, and $\eta$ is the measurement strength with the continuous weak measurement constraint, i.e., $\eta  \ll {\rm{1}}$. Also, $d{Y_t}$ is the independent and infinitesimal increment which represents the measurement output.

Generally, the unknown parameters may exist in the system Hamiltonian, the dissipation rates, or the measurement strength.
In this paper, we mainly focus on studying the estimation precision of single unknown parameter in the Hamiltonian.
The measurement process is assumed to be Markovian.
Due to the relationship between a normalized state ${\rho _t}$ and an un-normalized quantum state ${{\tilde \rho }_t}$, say ${\rho _t}{\rm{ = }}{{{{\tilde \rho }_t}}}/{{{\text{Tr}}\left( {{{\tilde \rho }_t}} \right)}}$ \cite{Gammelmark2013}, together with Eq.~\eqref{Eqe:stochastic_master_equation_1},
the increment $dY_t$ has the form
\begin{equation}\label{dY_t}
d{Y_t} = \sqrt \eta  Tr\left( {\mathcal{M}\left( {{\rho _t}} \right)} \right)dt + d{W_t},
\end{equation}
where $dW_t$ is the Wiener increment with zero mean and variance $dt$. The Eq.~\eqref{dY_t} describes the quantum fluctuations of the continuous output signal. Define ${\mathcal{L}_t} = \text{Tr}\left( {{{\tilde \rho }_t}} \right)$ as a  likelihood function. Owing to Eq.~\eqref{Eqe:stochastic_master_equation_1}, the derivative of the likelihood function ${\mathcal{L}_t}$ with respect to time $t$ can be written as \cite{Gammelmark2013,zhang2015precision}
\begin{equation}\label{dL_t}
\begin{aligned}
d{\mathcal{L}_t} &= \text{Tr}\left( {d{{\tilde \rho }_t}} \right) = \sqrt {\eta  } \text{Tr}\left( {\mathcal{M}\left( {{{\tilde \rho }_t}} \right)} \right)d{Y_t} \\
&= \sqrt {\eta  } \text{Tr}\left( {\mathcal{M}\left( {{\rho _t}} \right)} \right){\mathcal{L}_t}dY_t.
\end{aligned}
\end{equation}
Combining Eq.~\eqref{Eqe:stochastic_master_equation_1} with Eq.~\eqref{dL_t}, we can get the normalized quantum stochastic master equation by means of the multi-dimensional It\^{o} formula (one can refer to Appendix \ref{Mehtod:Ito} for details):
\begin{equation}\label{Eqe:stochastic_master_equation_2}
\begin{aligned}
d{\rho _t} =  - i\left[ {H,{\rho _t}} \right]dt + \eta {\cal D}\left[ F \right]{\rho _t}dt + \sqrt \eta  \mathcal{H}\left[ F \right]{\rho _t}d{W_t},
 \end{aligned}
\end{equation}
where $\mathcal{H}\left[ F \right]\rho  = {\cal M}\left( \rho  \right) - \rho \text{Tr}\left( {{\cal M}\left( \rho  \right)} \right).$
\section{Quantum parameter estimation in circuit QED}\label{Fisher_Information}
In this section, we propose an efficient algorithm to calculate the Fisher information by the measurement output and the likelihood function in circuit QED.
\subsection{The algorithm for calculting the Fisher information}\label{Algorithm}
Below, we use $l_t$ to denote the log-likelihood function, i.e., $l_t = \ln {\mathcal{L}_t}$ \cite{zhang2015precision,rossi2016enhanced}. From Eq.~\eqref{dL_t}, the derivative of $l_t$ with respect to time $t$ is described by
\begin{equation}\label{l_t}
 d{l_t} = d\ln {\mathcal{L}_t} = \frac{{d{\mathcal{L}_t}}}{{{\mathcal{L}_t}}} = \sqrt {\eta } \text{Tr}\left( {\mathcal{M}\left( {{\rho _t}} \right)} \right)d{Y_t}.
\end{equation}
Therefore, according to Eq.~\eqref{FI_0}, the Fisher information for single parameter estimation can be written as
\begin{equation}\label{FI}
I\left( \theta  \right) = E\left[ {{{\left( {\frac{{d\ln {\mathcal{L}_t}}}{{d\theta }}} \right)}^2}} \right] = E\left[ {{{\left( {\frac{{d{l_t}}}{{d\theta }}} \right)}^2}} \right].
\end{equation}
Substituting Eq.~\eqref{l_t} into Eq.~\eqref{FI}, we can obtain an analytic form of the Fisher information.
\begin{figure}[ht]
\centering
\includegraphics[width=7.5cm]{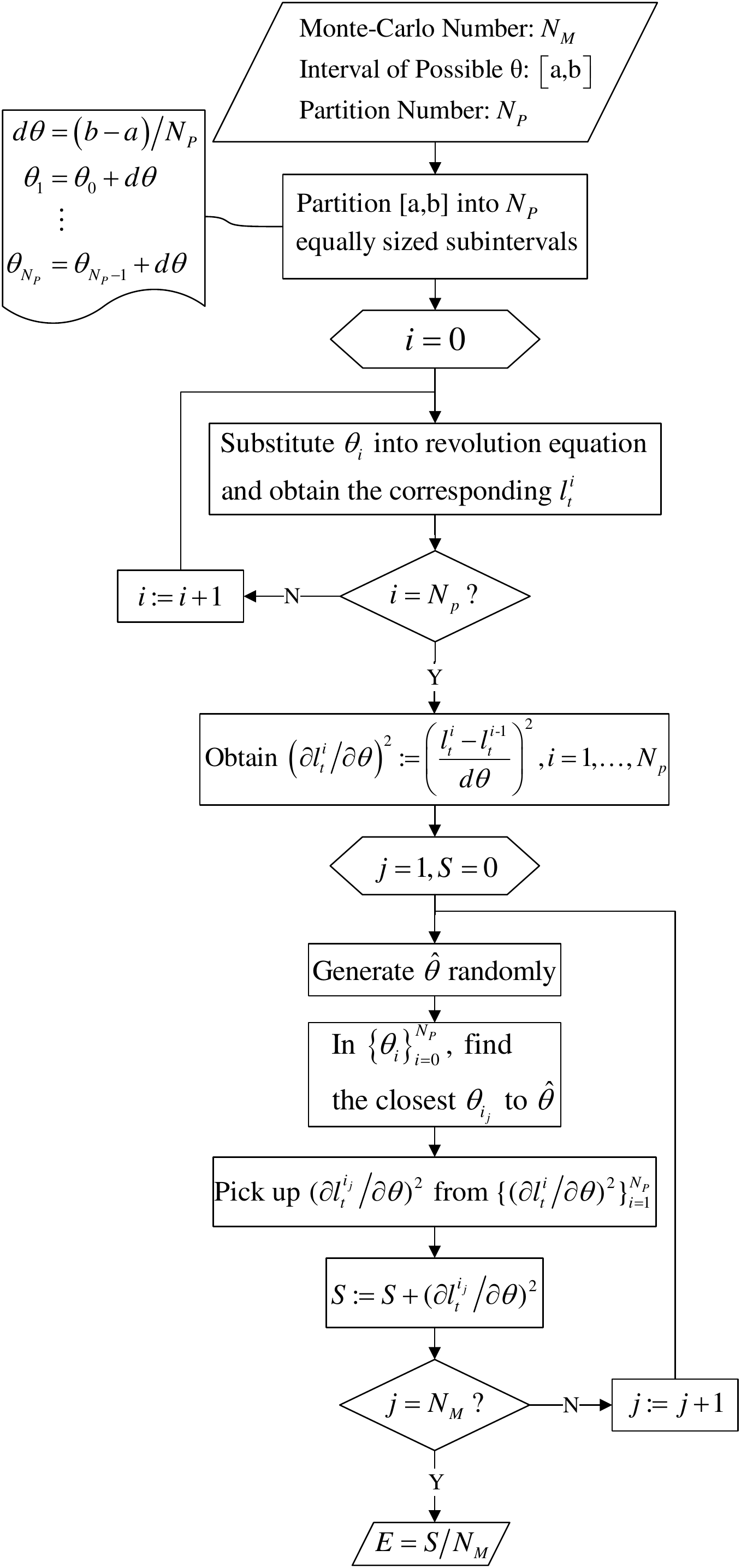}
\caption{The procedure of calculating the Fisher information via dispersive measurement.}
\label{fig:processFI}
\end{figure}

From the Fisher information Eq.~\eqref{FI}, it is easy to find that $\theta$ is not an independent variable of the likelihood function.
In other words, there do not exist an explicit expression of $l_t$ with respect to $\theta$, which makes the calculation of $I(\theta)$ remarkable difficulty.
In order to efficiently calculate the Fisher information, we propose a numerical algorithm with the help of the MH algorithm \cite{Gilks1996Introducing} and the MCMC integration \cite{Efendiev:2008}.

In the beginning, we set a series of the unknown parameter $\left\{ {{\theta _i}} \right\}$ satisfying
\begin{equation}\label{theta}
{{\theta _{i + 1}} = {\theta _i} + d\theta ,~~i = 0,1,2,\dots,N_P,}
\end{equation}
where the interval $d\theta$ is a small constant.
For each $\theta _i$, there exists a log-likelihood function, say $l^i _t$, corresponding to $\theta _i$.
Here, the collection of log-likelihood functions { $\{ {l_t^0,l_t^1, \ldots ,l_t^{{N_P}}} \}$} is a set of functions of time with $t \in \left[ {0,T} \right]$.
As we can see in {Eq.~\eqref{FI}}, calculating the Fisher information requires to calculate the derivative of $l _t$ with respect to $\theta$ firstly.
However, the noise induced by the measurement process makes it improper to use the ordinary numerical derivation to compute $\{{dl_t^i/d\theta}\}$.
To deal with it, it is natural to use the average evolution of the ensemble to eliminate the impact of measurement noise.
Since $d\theta$ can be infinitesimal, the  derivative of $l _t$ with respect to $\theta$ can be given by the {Newton's backward difference quotient} with infinitesimal errors, i.e.,
\begin{equation}\label{derivation_l_t}
{\frac{{d{l^i_t}}}{{d\theta }} \approx \frac{{l_t^{i} - l_t^{i-1}}}{{{\theta _{i }} - {\theta _{i-1}}}} = \frac{{l_t^{i} - l_t^{i-1}}}{{d\theta }},i = 1,2, \ldots ,{N_P}.}
\end{equation}
Next, we randomly generate a cluster of $\theta$ by the MH algorithm (one can refer to the Appendix \ref{Mehtod:Metropolis} for details), whose prior probability distribution is assumed to satisfy a certain distribution.
Denote such generated cluster of $ \theta$  by
\begin{equation}\label{hat_theta}
\hat \theta  = \left\{ {{{\hat \theta }_j}\left| {j = 1,2,\dots,N_M} \right.} \right\},
\end{equation}
where $N_M$ is the Monte Carlo number. Note that the number of candidate points  $N_A $ that used to generate random samples is chosen to be larger than the Monte Carlo number, i.e, ${N_M} \le {N_A}$. In the set of $\hat\theta$, the fluctuation of the pre-estimated parameter values is rather small. This process makes the following numerical calculation as close to the analytic result as possible.
For simplicity, one may anticipate the initial value of the sequence generating $\hat\theta$ to be a constant value. It is easy to choose the closest $\theta_{i_j}$  to each $\hat\theta_j$ by comparing $\hat\theta$ with $\theta$.
As a result, $\{(dl_t^{i_1}/d\theta)^2,(dl_t^{i_2}/d\theta)^2,\dots,(dl_t^{i_{N_M}}/d\theta)^2\}$ could be picked out from the collection {$\{(dl_t^{1}/d\theta)^2, \dots, (dl_t^{{N_P}}/d\theta)^2\}$ }determined by the generated $\hat\theta$.
\begin{figure}[ht]
\centering
\includegraphics[width=8.5cm,height=4.5cm]{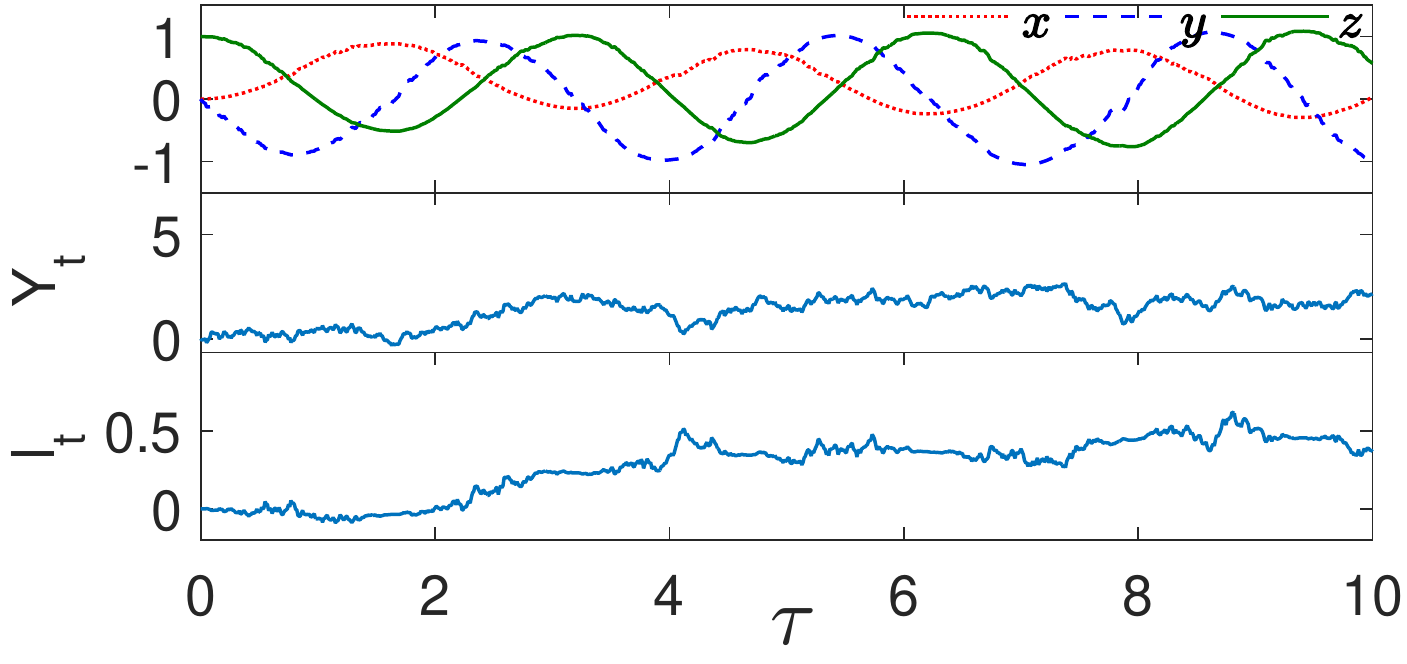}
\caption{(Color online) The top panel shows the evolution of the three components of the Bloch vector with $\Omega=1, \Delta=1.73$ and $\eta = 0.01$. The curves are $x = \text{Tr}\left( {{\sigma _x}{\rho _t}} \right)$ (the dotted red curve), $y = \text{Tr}\left( {{\sigma _y}{\rho _t}} \right)$ (the dashed blue curve), and $z = \text{Tr}\left( {{\sigma _z}{\rho _t}} \right)$ (the soiled green curve).
The middle one represents the measurement output $Y_t$. The bottom panel shows the log-likelihood function $l_t$.}
\label{fig:one_measurement}
\end{figure}

Finally, calculating  the Fisher information means to acquire the expected value $E[(d{l_t}/d\theta)^2]$ from the sample $\{(dl_t^{i_1}/d\theta)^2, (dl_t^{i_2}/d\theta)^2, \dots, (dl_t^{i_{N_M}}/d\theta)^2\}$ owing to Eq.~\eqref{FI}. By the Makov chain Monte Carlo integration, see Appendix \ref{Mehtod:Makov} for details, the Fisher information can be approximated as
\begin{equation}\label{FIV}
E\left[ {{{\left( {{{d {l_t}} \mathord{\left/
{\vphantom {{d {l_t}} {d \theta }}} \right.
\kern-\nulldelimiterspace} {d \theta }}} \right)}^2}} \right] \approx \frac{1}{{{N_M}}}{\sum\limits_{j = 1}^{{N_M}} {\left( {\frac{{d l_t^{i_j}}}{{d \theta }}} \right)} ^2}.
\end{equation}
As a conclusion, the procedure of calculating the Fisher information is shown in Figure.~\ref{fig:processFI}.
\subsection{Numerical simulations}\label{Simulation}
Let $\Omega$ in the Hamiltonian \eqref{H} be an unknown parameter that requires estimating. We denote the normalized quantum  state $\rho_t$ by
\begin{equation}\label{rho_0}
 {\rho _t} = \frac{1}{2}\left( {\begin{array}{*{20}{c}}
{1 + z}&{x - iy}\\
{x + {\rm{iy}}}&{1 - z}
\end{array}} \right),
\end{equation}
and the initial state is $\rho _0 =1/2\left| {{\psi _0}} \right\rangle \langle {\psi _0}|$ with
$| {{\psi _{\rm{0}}}}\rangle=({\begin{matrix}1 & 0 \end{matrix}})^T$, i.e.,
$x(0) = y(0) = 0,~z(0) = 1$.
The other parameters in the stochastic master equation \eqref{Eqe:stochastic_master_equation_2} are $\Delta = 1.73$ and $\eta  =0.01$, and the measurement operator is given by $F=\sigma _y$.
For convenience, we define $\tau=\Omega t$ throughout this section.
Suppose that an initial reference value of the unknown parameter $\Omega$ is set to be $1$, then
the sequence $\hat\Omega$ can be obtained by proceeding the MH algorithm when the stationary distribution and proposal distribution are assumed to satisfy the normal distributions $N(0,1)$ and $N(0,dt)$, respectively.
Fig.~\ref{fig:one_measurement} shows the evolution of the normalized quantum state with dispersive measurement according to Eq.~\eqref{Eqe:stochastic_master_equation_2}.
The output $Y_t$ and the log-likelihood function $l_t$ are also plotted.
\begin{figure}[ht]
\begin{minipage}[h]{0.49\linewidth}
\centerline{\includegraphics[width=4.4cm]{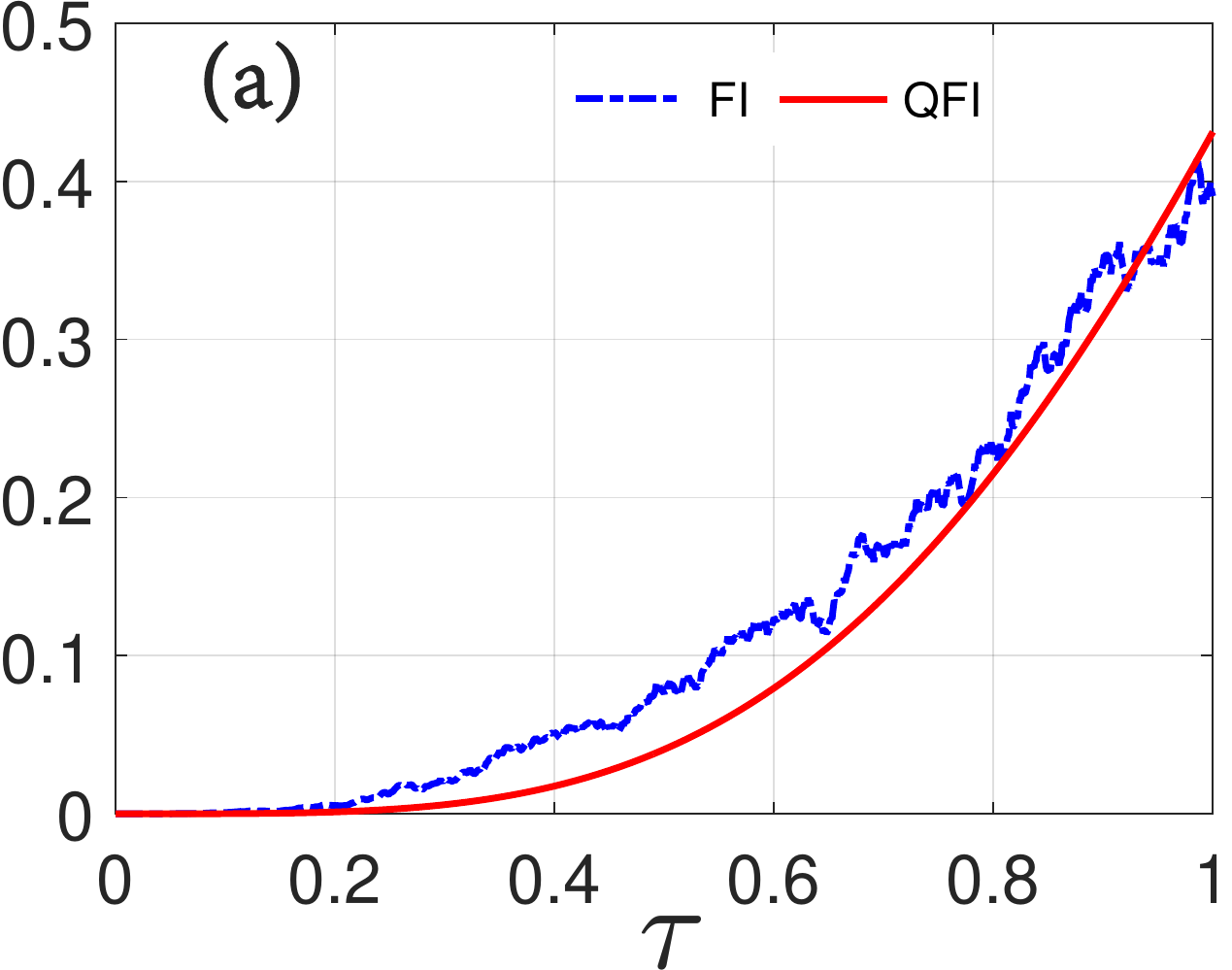}}
\end{minipage}
\begin{minipage}[h]{0.49\linewidth}
\centerline{\includegraphics[width=4.4cm]{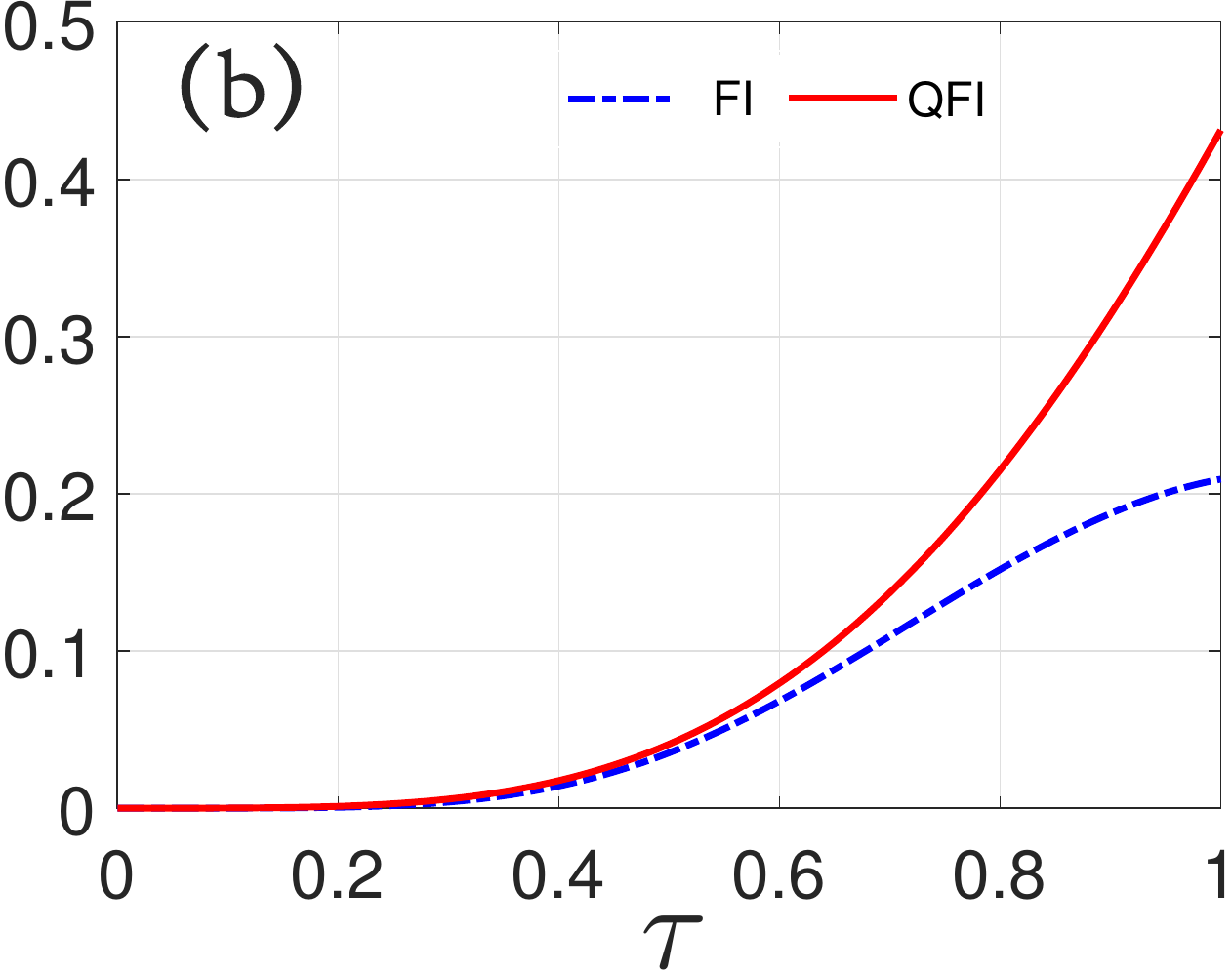}}
\end{minipage}
\caption{(Color online) The blue dash-dotted curves in (a) and (b) represent the single evolution and ensemble evolution  of the Fisher information, respectively. The red solid curves are the evolution of the quantum Fisher information.
}\label{fig:measurements}
\end{figure}

Based on the proposed algorithm and the stochastic master equation (8), we show the evolution of the Fisher information for quantum parameter estimation in Fig.~(3).  The blue dash-dotted curve in Fig.~3(a) represents the single evolution of the Fisher information, and the blue dash-dotted curve in Fig.~3(b) is the ensemble evolution with $500$ dispersive measurements in circuit QED.  The red solid curves in Fig. 3 are the evolution of the quantum Fisher information with the help of the definition, Eq.~(3).
The quantum Fisher information always represents the upper bound of the Fisher information. From Fig.~(3), we find that the single evolution of the Fisher information can probably approach the quantum Fisher information. The same phenomenon is observed in the ensemble evolution in the short time interval. These results demonstrate the effectiveness of the proposed algorithm. 

Furthermore, we plot the ensemble evolutions of the Fisher information with the proposed algorithm for various measurement operators in Fig.~(4). The green dashed, blue dot-dashed and purple dotted curves are the $\sigma_x$, $\sigma_y$ and $\sigma_z$ measurements in circuit QED, respectively.  According to the original definition,   quantum Fisher information is the Fisher information that optimized over all possible measurement operators allowed by quantum mechanics. Searching the optimal measurement  operator remains to be further studied. 

\begin{figure}
  \centering
  \includegraphics[width=8cm,height=6cm]{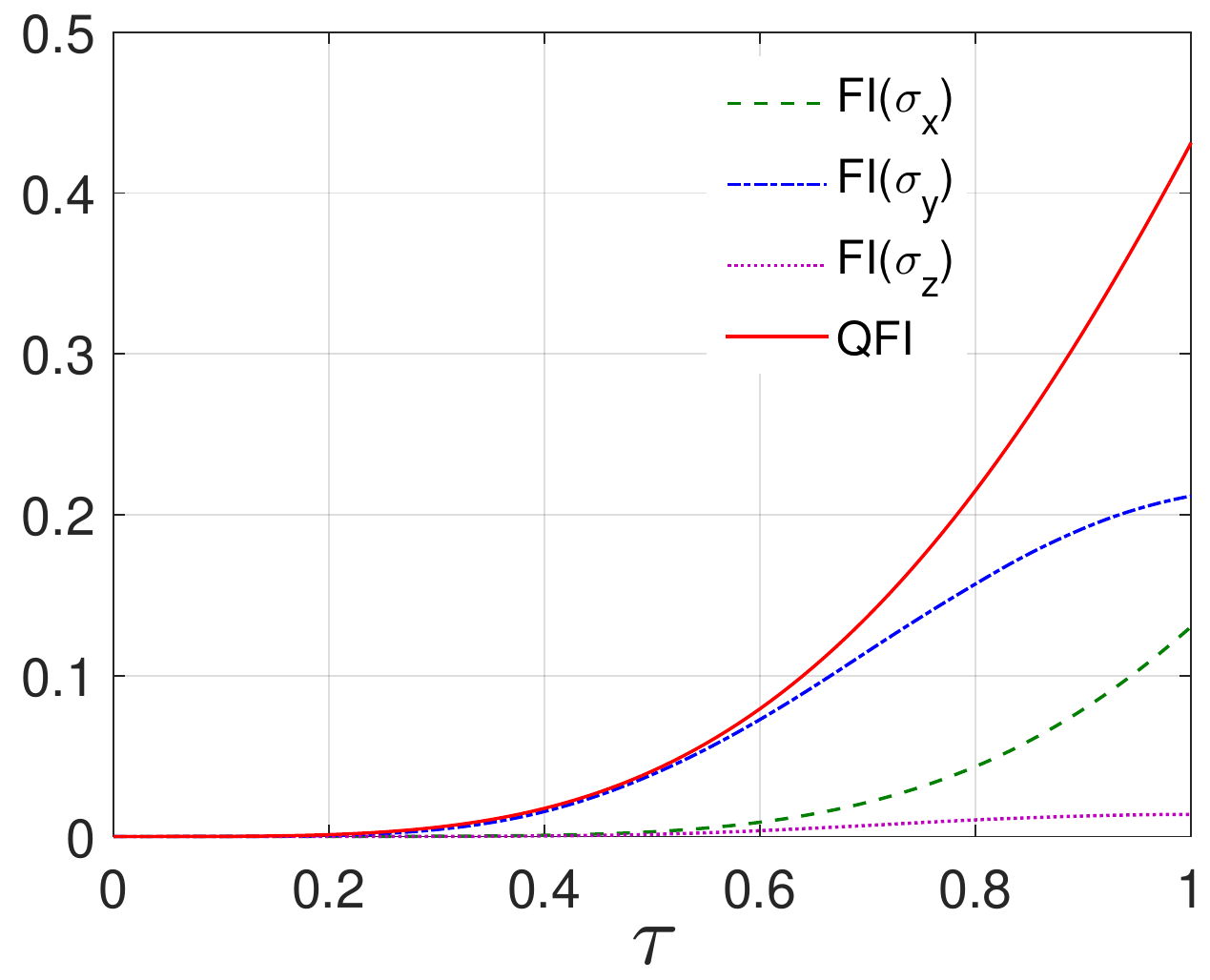}
  \caption{(Color online) The ensemble evolutions of the Fisher information with the proposed algorithm for various measurement operators. }\label{fig:compare}
\end{figure}

\section{Conclusion}\label{Conclusion}
We discussed the quantum parameter estimation in circuit QED via dispersive measurement and the stochastic master equation.
Based on the Metropolis Hastings algorithm and the Markov chain Monte Carlo integration, a new algorithm is proposed to calculate the Fisher information.
Numerical results demonstrate that the single evolution of the Fisher information can probably approach the quantum Fisher information. The same phenomenon is observed in the ensemble evolution in the short time interval. Finally, we discussed the ensemble evolutions of the Fisher information with the proposed algorithm for various measurement operators.

\section*{Acknowledgements}
This work is mainly supported by the National Natural Science Foundation of China under Grant 11404113, and the Guangzhou Key Laboratory of Brain Computer Interaction and Applications under Grant 201509010006.

\appendix
\section*{Appendix}
\subsection{The lemma of the multi-dimensional It\^{o} formula}\label{Mehtod:Ito}
In the multi-dimensional It\^{o} formula, it's worth noting that if $x(t)$ were continuously differentiable with respect to time $t$, then the term $\frac{1}{2}d{x^T}(t){V_{xx}}({x(t),t})dx(t)$ would not appear owing to the classical calculus formula for total derivatives. For example, if $V( {{x_1},{x_2}})$ is continuously differentiable  with respect to $t$, e.g., $V( {{x_1},{x_2}}) = {x_1}( t){x_2}(t)$, then it's derivation should be $dV( {{x_1},{x_2}}) = {x_1}d{x_2} + {x_2}d{x_1} + d{x_1}d{x_2}$.
\begin{figure}[ht]
\centering
\includegraphics[width=8.5cm]{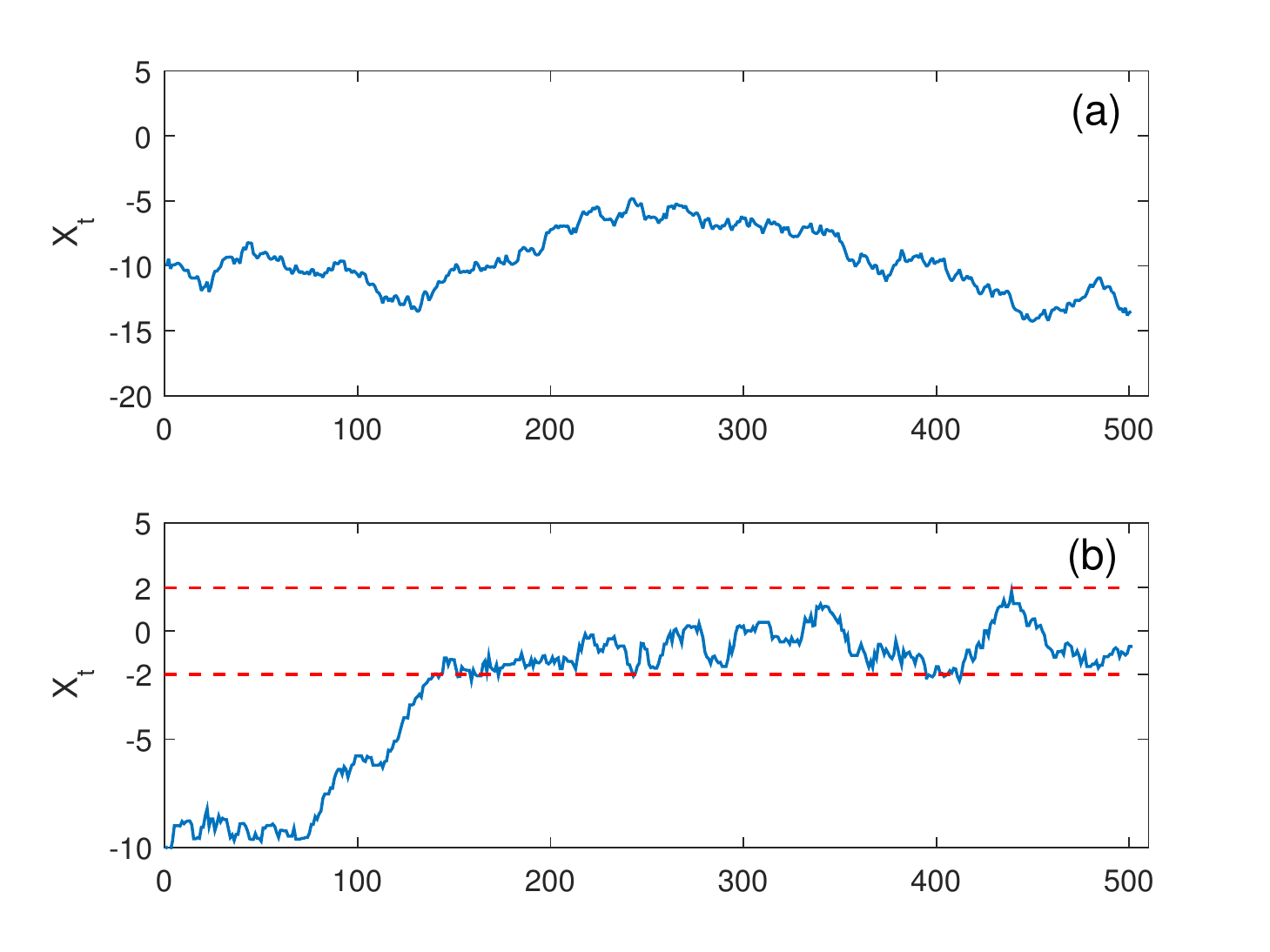}
\caption {(Color online) We illustrate the sampling process of the Metropolis Hastings algorithm. Here, the initial value is $X(1) = -10$.
Fig.~5(a) represents the stationary  distribution $N(0,0.1)$.
In Fig.~5(b), we plot 500 iterations from Metropolis Hastings algorithm with the stationary distribution $N(0,1)$ and proposal distribution $N(0,0.1)$.
Obviously,  sampling data selecting from the  latter part would be better.}
\label{fig:example}
\end{figure}
\subsection{Metropolis Hastings algorithm \cite{Gilks1996Introducing} }\label{Mehtod:Metropolis}
In Markov chains, suppose we generate a sequence of random variables ${X_1},{X_2},...,{X_{n }}$ with Markov property, namely the probability of moving to the next state depends only on the present state and not on the previous state:
\begin{equation} \nonumber
\begin{aligned}
&\Pr \left\{ {{X_{n + 1}} = x\left| {{X_1} = {x_1},...,{X_n} = {x_n}} \right.} \right\} \\
&~~~= \Pr \left\{ {{X_{n + 1}} = x\left| {{X_n} = {x_n}} \right.} \right\}.
\end{aligned}
\end{equation}
Then, for a given state $X_t$, the next state $X_{t+1}$ does not depend further on the hist of the chain ${X_1},{X_2},...,{X_{t - 1}}$, but comes from a distribution which only on the current state of the chain $X_t$. For  any time instant $t$, if the next state is the first sample reference point $Y$ obeying distribution $q\left( { \cdot \left| {{X _t}} \right.} \right)$ which is called the transition kernel of the chain, then obviously it depends on the current state $X_t$. In generally,  $q\left( { \cdot \left| {{X _t}} \right.} \right)$ may be a multidimensional normal distribution with mean $X$, so the candidate point $Y$ is accepted with probability $\alpha \left( {{X _t},Y} \right)$ where
\begin{equation*}
\alpha \left( {X ,Y} \right) = \min \left( {1,\frac{{\pi \left( Y \right)q\left( {X \left| Y \right.} \right)}}{{\pi \left( X  \right)q\left( {Y\left| X  \right.} \right)}}} \right).
\end{equation*}
Here, $\pi \left( A \right)$ stands a function only depends on $A$.
If the candidate point is accepted, the next state becomes $X_{t+1}=Y$. If the candidate point is rejected, it means that the chain does not move, the next state will be $X_{t+1}=X$. We illustrate this sampling process with a simple example, see Fig.~\ref{fig:example}.
Here, the initial value is $X(1) = -10$.
Fig.~\ref{fig:example}(a) represents the stationary distribution $N(0,0.1)$.
In Fig.~\ref{fig:example}(b), we plot 500 iterations from Metropolis Hastings algorithm with the stationary distribution $N(0,1)$ and proposal distribution $N(0,0.1)$.
Obviously, sampling data selecting from the latter part would be better.
\subsection{Makov Chain Monte Carlo integration \cite{Efendiev:2008}}\label{Mehtod:Makov}
In Markov chain, the Monte Carlo integration can be used to evaluate $E[ f( X )]$ by drawing samples $\{ {X _1},...{X _n} \}$ from the Metropolis Hastings algorithm. Here
\begin{equation*}
E\left[ {f\left( X  \right)} \right] \approx \frac{1}{n}\sum\limits_{i = 1}^n {f\left( {{X _i}} \right)},
\end{equation*}
means that the population mean of $f\left(X \right)$ is approximated by the sample mean.
When the sample ${X_t}$ are independent, the law of large numbers ensures that the approximation can be made as accurate as desired by increasing the sample. Note that here $n$ is not the total amount of samples by Metropolis Hastings algorithm but the length of drawing samples.
\label{sec:TeXbooks}


\begin{thebibliography}{10}




\bibitem{Helstrom1976}
C.~W.~Helstrom, Quantum detection and estimation theory, J. Statist. Phy. \textbf{1}, 231-252 (1969).

\bibitem{Holevo2011}
A.~Holevo, \emph{Probabilistic and Statistical Aspects of Quantum Theory}, (Edizioni della Normale, Pisa, 2011).

\bibitem{Wiseman2010}
H.~M.~Wiseman and G.~J.~Milburn, \emph{Quantum Measurements and Control}, (Cambridge University Press, Cambridge, 2010).


\bibitem{Escher2011}
B.~M.~Escher, R.~L.~de~Matos~Filho, and	L.~Davidovich, General framework for estimating the ultimate precision limit in noisy quantum-enhanced metrology, Nature Physics \textbf{7}, 406-411 (2011).

\bibitem{giovannetti:2004}
V.~Giovannetti, S.~Lloyd, and L.~Maccone, Quantum-enhanced measurements: beating the standard quantum limit, Science \textbf{306}, 1330-1336 (2004).

\bibitem{caves:1980}
C.~M.~Caves, K.~S.~Thorne, R.~W.~Drever, V.~D.~Sandberg, and M.~Zimmermann, On the measurement of a weak classical force coupled to aquantum-mechanical oscillator.~i.~issues of principle, Rev. Mod. Phys. \textbf{52}, 341-392 (1980).

\bibitem{caves:1981}
C.~M.~Caves, Quantum-mechanical noise in an interferometer, Phys. Rev. D \textbf{23}, 1693-1708 (1981).

\bibitem{zwierz:2010}
M.~Zwierz, C.~A.~P{\'e}rez-Delgado, and P.~Kok, General optimality of the Heisenberg limit for quantum metrology, Phys. Rev. Lett. \textbf{105} 180402 (2010).

\bibitem{hodges2012cramer}
J. L.~Hodges and E.~L.~Lehmann, \emph{Some Applications of the Cram\'{e}r-Rao Inequality}, (Springer, Boston, 2012).

\bibitem{Zhong:2013}
W.~Zhong, Z.~Sun, J.~Ma, X.~G.~Wang, and F.~Nori, Fisher information under decoherence in Bloch representation, Phy. Rev. A \textbf{87}, 022337 (2013).

\bibitem{Wang:2016}
Z.~H.~Wang, Q.~Zheng, X.~Wang, and Y.~Li, The energy-level crossing behavior and quantum Fisher information in a quantum well with spin-orbit coupling, Sci. Rep. \textbf{6}, 22347 (2016).


\bibitem{Li:2013}
N.~Li and S.~Luo, Entanglement detection via quantum Fisher information, Phys. Rev. A \textbf{88}, 014301 (2013).


\bibitem{Zhang:2013}
Y.~Zhang, X.~W.~Li, W.~Yang, and G.~R.~Jin, Quantum Fisher information of entangled coherent states in the presence of photon loss, Phys. Rev. A \textbf{88}, 043832 (2013).

\bibitem{Jacobs2014Quantum}
K.~Jacobs, \emph{Quantum Measurement Theory and Its Applications}, (Cambridge University Press, Cambridge, 2014).


\bibitem{Smith:2006}
G.~A.~Smith, A.~Silberfarb, I.~H.~Deutsch, and P.~S.~Jessen, Efficient quantum state estimation by continuous weak measurement and dynamical control, Phys. Rev. Lett. \textbf{97}, 180403 (2006).

\bibitem{aharonov1988result}
Y.~Aharonov, D.~Z.~Albert, and L.~Vaidman, How the result of a measurement of a component of the spin of a spin-1/2 particle can turn out to be 100, Phys. Rev. Lett. \textbf{60}, 1351-1354 (1988).

\bibitem{zhang2015precision}
L.~Zhang, A.~Datta, and I.~A.~Walmsley, Precision metrology using weak measurements, Phys. Rev. Lett. \textbf{114}, 210801 (2015).


\bibitem{Xu:2013}
X.~Y.~Xu, Y.~Kedem, K.~Sun, L.~Vaidman, C.~F.~Li, and G.~C.~Guo, Phase estimation with weak measurement using a white light source, Phys. Rev. Lett. \textbf{111}, 033604 (2013).

\bibitem{Breuer:2002}
H.~P.~Breuer and F.~Petruccione, \emph{The Theory of Open Quantum Systems}, (Oxford University Press, New York, 2002).

\bibitem{Ralph2011}
J.~F.~Ralph, K.~Jacobs, and C.~D.~Hill, Frequency tracking and parameter estimation for robust quantum state estimation, Phys. Rev. A \textbf{84}, 052119 (2011).


\bibitem{Gammelmark2013}
S.~Gammelmark and K.~M\o lmer, Bayesian parameter inference from continuously monitored quantum systems, Phys. Rev. A \textbf{87}, 032115 (2013).

\bibitem{Gammelmark2014}
S.~Gammelmark and K.~M\o lmer, Fisher information and the quantum Cram\'{e}r-Rao sensitivity limit of continuous measurements, Phys. Rev. Lett \textbf{112}, 170401 (2014).

\bibitem{You:2005}
J.~Q.~You and F.~Nori, Superconducting circuits and quantum information, Physics Today \textbf{58}(11), 42-47 (2005).

\bibitem{You:2011}
J.~Q.~You and F.~Nori, Atomic physics and quantum optics using superconducting circuits, Nature \textbf{474}, 589-597 (2011).

\bibitem{Genoni:2017}
M.~G.~Genoni, Cram\'{e}r-Rao bound for time-continuous measurements in linear Gaussian quantum systems, Phys. Rev. A \textbf{95}, 012116 (2017).

\bibitem{Blais:2004}
A.~Blais, R.~S.~Huang, A.~Wallraff, S.~M.~Girvin, and R.~J.~Schoelkopf, Cavity quantum electrodynamics for superconducting electrical circuits: an architecture for quantum computation, Phys. Rev. A \textbf{69}, 062320 (2004).

\bibitem{Wallraff:2004}
A.~Wallraff, D.~I.~Schuster, A.~Blais, L.~Frunzio, R.~S. Huang, J.~Majer, S.~Kumar, S.~M.~Girvin, and R.~J.~Schoelkopf, Strong coupling of a single photon to a superconducting qubit using circuit quantum electrodynamics, Nature \textbf{431}, 162-167 (2004).

\bibitem{Chiorescu:2004}
I.~Chiorescu, P.~Bertet, K.~Semba, Y.~Nakamura, C.~J.~P.~M.~Harmans, and J.~E.~Mooij, Coherent dynamics of a flux qubit coupled to a harmonic oscillator, Nature \textbf{431}, 159-162 (2004).

\bibitem{Vijay:2012}
R.~Vijay, C.~Macklin, D.~H.~Slichter, S.~J.~Weber, K.~W.~Murch, R.~Naik, A.~N.~Korotkov, and I.~Siddiqi, Stabilizing Rabi oscillations in a superconducting qubit using quantum feedback, Nature \textbf{490}, 77-80 (2012).


\bibitem{Slichte:2012}
D.~H.~Slichter, R.~Vijay, S.~J.~Weber, S.~Boutin, M.~Boissonneault, J.~M.~Gambetta, A.~Blais, and I.~Siddiqi, Measurement-induced qubit state mixing in circuit QED from up-converted dephasing noise, Phys. Rev. Lett. \textbf{109}, 153601 (2012).

\bibitem{Xiang:2013}
Z.~L.~Xiang, S.~Ashhab, J.~Q.~You, and F.~Nori, Hybrid quantum circuit consisting of a superconducting flux qubit coupled to both a spin ensemble and a transmission-line resonator, Phys. Rev. B \textbf{87}, 144516 (2013).

\bibitem{Cui:2013}
W.~Cui and F.~Nori, Feedback control of Rabi oscillations in circuit QED, Phys. Rev. A \textbf{88}, 063823 (2013).


\bibitem{Gilks1996Introducing}
W.~R.~Gilks, S.~Richardson, and D.~J.~Spiegelhalter, \emph{Markov Chain Monte Carlo in Practice}, (Chapman \& Hall, London, 1996).

\bibitem{Johnston2014Efficient}
I.~G.~Johnston, Efficient parametric inference for stochastic biological systems with measured variability, Stat. Appl. Genet. Mol. Biol. \textbf{13}, 379-390 (2014).


\bibitem{Efendiev:2008}
Y.~Efendiev, A.~Datta-Gupta, X.~Ma, and B.~Mallick, Modified markov chain monte carlo method for dynamic data integration using streamline approach, Math. Geosci. \textbf{40}, 213-232 (2008).


\bibitem{Braunstein1994Statistical}
S.~L.~Braunstein and C.~M.~Caves, Statistical distance and the geometry of quantum states, Phys. Rev. Lett. \textbf{72}(22), 3439-3443 (1994).

\bibitem{Ciampini:2016}
M.~A.~Ciampini, N.~Spagnolo, C.~Vitelli, L.~Pezz\`{e}, A.~Smerzi, and F.~Sciarrino, Quantum-enhanced multiparameter estimation in multiarm interferometers, Sci. Rep. \textbf{6}, 28881 (2016).

\bibitem{Kiilerich2016}
A.~H.~Kiilerich and K.~M\o lmer, Bayesian parameter estimation by continuous homodyne detection, Phys. Rev. A \textbf{94}, 032103 (2016).

\bibitem{rossi2016enhanced}
M.~A.~C.~Rossi, F.~Albarelli, and M.~G.~A.~Paris, Enhanced estimation of loss in the presence of kerr nonlinearity, Phys. Rev. A \textbf{93}, 053805 (2016).

\bibitem{Devoret:2013}
M.~H.~Devoret and J.~R.~Schoelkopf, Superconducting circuits and quantum information: an outlook, Science \textbf{339}, 1169-1174 (2013).

\bibitem{Gambetta:2008}
J.~Gambetta, A.~Blais, M.~Boissonneault, A.~A.~Houck, D.~I.~Schuster, and S.~M.~Girvin, Quantum trajectory approach to circuit QED: Quantum jumps and the Zeno effect, Phy. Rev. A \textbf{77}, 012112 (2008).

\bibitem{Qi:2010}
B.~Qi and L.~Guo, Is measurement-based feedback still better for quantum control systems? System Control Letters \textbf{59}, 333-339 (2010).

\bibitem{Feng:2016}
W.~Feng, P.~F.~Liang, L.~P.~Qin, and X.~Q.~Li, Exact quantum Bayesian rule for qubit measurement in circuit QED, Sci. Rep. \textbf{6}, 20492 (2016).



\end{thebibliography}
\end{document}